\documentclass[pra,aps,twocolumn,showpacs,10pt]{revtex4-1}

\usepackage{amsmath,amssymb,graphicx,amsthm,dsfont,bm,color,ulem,adjustbox}

\begin{document}

\title{Optical vortex trapping and annihilation by means of nonlinear Bessel beams in nonlinearly absorbing media }

\author{Jos\'e L. Garc\'{i}a-Riquelme}
\affiliation{Grupo de Sistemas Complejos, ETSIME, Universidad Polit\'ecnica de Madrid, Rios Rosas 21, 28003 Madrid, Spain}
\author{Francisco Ramos}
\affiliation{Nanophotonics Technology Center, Universitat Polit$\grave{\textit{e}}$cnica de Val$\grave{\textit{e}}$ncia, Camino de Vera s/n, 46022 Valencia, Spain}
\author{Miguel A. Porras}
\affiliation{Grupo de Sistemas Complejos, ETSIME, Universidad Polit\'ecnica de Madrid, Rios Rosas 21, 28003 Madrid, Spain}

%\affil[*]{Corresponding author: miguelangel.porras@upm.es}

%\ociscodes{(050.1940) Diffraction; (140.3300) Laser beam shaping; (320.5550) Pulses; (070.7345) Paraxial wave optics.}

%\doi{\url{http://dx.doi.org/10.1364/ao.XX.XXXXXX}}

\begin{abstract}
In nonlinear Kerr media at intensities such that multiphoton absorption is significant, a vortex of topological charge $m$ in the center of a high-order nonlinear Bessel beam (NBB) can be stable and subsist endlessly. We show that the $m$-charged NBB is not only stable but is formed spontaneously from any other $n$-charged NBB and $N$ ``foreign" vortices of total charge $s$ randomly nested in the beam cross section if $n+s=m$. All nested vortices merge in the center of the original NBB, which undergoes a mode conversion to the NBB that preserves the topological charge and the inward-directed power current that sustains the diffraction-free and attenuation-free propagation in the medium with nonlinear absorption. We foresee different applications such as the creation of stable, multiply charged vortices without tight alignment requirements but by spontaneous vortex combination, mixing waves or particles that the vortices can guide, fast annihilation of vortex dipoles, and cleaning of speckled beams by massive annihilation of vortices.
\end{abstract}

\maketitle

%\noindent Introduction

\section{Introduction}

The dynamics of optical vortices in optical fields has been a matter of interest in the last decades, during which the fundamental laws that govern this dynamics have been established and applied to steer the motion of vortices and vortex arrays embedded in different background fields in diverse media \cite{ROZAS,KIVSHAR,KIVSHAR2,LUTHER,VBESSEL,ROZAS2,ROZAS3,HASINGER,PORRAS3}. Stabilizing the vortices and controlling their motion is of utmost importance in diverse applications such as waveguide writing, wave mixing and particle trapping \cite{CID,SALGUEIRO,CARLSSON,SHEDOV}.
In other situations, the presence of vortices in a light beam is highly undesirable, such as the vortices appearing spontaneously in speckled fields during propagation in a turbulent atmosphere, since they strongly deteriorate the optical performance of the light beam \cite{RUX,VLADIMIROVA,WANG,RUX2}. Considerable effort has been paid to reduce the speckle of scintillated beams using different techniques such as diffractive optical elements \cite{WANG} or adaptive optics \cite{RUX2}.

Recently we have proposed nesting optical vortices in the diffraction-free, nonlinear Bessel beams (NBBs) supported by transparent self-defocusing media as a way to prolong the distance at which the vortices survive stably, at the same time that their motion mimics the simple motion of vortices in an uniform, plane wave background, and is therefore easily predictable \cite{PORRAS3}. The only shortcoming of this proposal is the small, but intrinsic instability of NBB background in transparent self-defocusing media. The development of the instability triggered by noise or any small imperfection leads to the destruction of the NBB-vortex system, although at distances one order of magnitude larger than typical distances of vortex decay in standard Gaussian-like backgrounds.

In this paper we describe the dynamics and applications of vortices embedded in the diffraction-free NBBs supported by self-defocusing or self-focusing media at intensities at which nonlinear absorption due to multiphoton absorption is a relevant mechanism of energy dissipation. The existence of diffraction-free and attenuation-free, fundamental (vortex-free) NBBs in Kerr media with multiphoton absorption is known about one decade ago \cite{PORRAS1}. These fundamental NBBs have been shown to play a prominent role in light filaments excited by Bessel beams \cite{POLESANA1}, in which case multiphoton absorption produces ionization. High-order NBBs carrying a vortex of arbitrary topological charge in their center have also been described recently \cite{RUIZ,JUKNA}, and have subsequently been realized experimentally to induce tubular filamentation for material processing \cite{XIE}.

In contrast to NBBs in transparent media \cite{PORRAS3,JOHANNISSON}, the fundamental and high-order NBBs in media with multiphoton absorption can be completely stable against perturbations \cite{PORRAS2}. In relation to ``foreign" vortices embedded in a fundamental or high-order NBB, we have found that their dynamics is extremely simple and predictable: If the NBB background is robust enough to survive the perturbation produced by the nested vortices, all them, regardless of their initial positions in the NBB cross section are directed towards the NBB center, where they combine at the same time that the NBB experiences a conversion to the NBB that preserves the total topological charge. The NBB center is therefore a stable equilibrium point for vortices, where they remain at rest endlessly once they have been trapped. We explain analytically the vortex attraction property of NBBs in media with multiphoton absorption based on their particular feature of being propagation-invariant with permanently converging wave fronts \cite{PORRAS1}, and on the well-known laws governing the dynamics of vortices in a phase gradient \cite{KIVSHAR,KIVSHAR2}.

This dynamics is in sharp contrast with the dynamics of vortices in standard, Gaussian-like, or super-Gaussian backgrounds, where the vortices usually spiral out and decay in the divergent wave fronts upon diffraction \cite{ROZAS,ROZAS2,ROZAS3}. This opposite dynamics suggests quite different applications. We demonstrate numerically vortex trapping that allows to create vortex beams of arbitrary topological charge without any alignment requirements \cite{ZUKAUSKAS}, but just by inserting a vortex anywhere in the fundamental Bessel beam. Also, inserting a set of vortices at different positions results in the spontaneous formation of a high-order NBB of the total topological charge, a property that can be used to create a $N$ to $1$ combiner of particles \cite{SHEDOV} or of the vortex-guided waves for wave mixing \cite{CID,SALGUEIRO,CARLSSON}. In particular, the two vortices of a vortex dipole seldom annihilate, but move along straight parallel trajectories in an infinite plane wave \cite{KIVSHAR}, and may annihilate or not in a Gaussian beam \cite{RUX}. In a NBB they annihilate spontaneously at a short propagation distance. Based on this property, we demonstrate multiple annihilation of many singly-charged, positive and negative vortices contained by speckled beams by simply passing the beam though an axicon and a nonlinear absorbing, self-focusing or self-defocusing medium occupying the Bessel zone after the axicon.

\section{Nonlinear Bessel Beams in Kerr media with multiphoton absorption}\label{NBB}

In this section we recall the properties of the fundamental \cite{PORRAS1} and high-order NBBs \cite{RUIZ,JUKNA}, stressing those that explain their vortex attractor property. Nonlinear Bessel beams are well-known in self-focusing Kerr media; here we extend the description to self-defocusing media. Indeed, their existence does not critically depend on the dispersive nonlinearities, so the inclusion of other nonlinearities such as Kerr saturation does not substantially alter the NBB properties \cite{RUIZ}. Nonlinear absorption is also optional \cite{JOHANNISSON} for the existence of NBBs, but it is crucial for their stability \cite{PORRAS2}. Nonlinear absorption due to multiphoton absorption of different orders arises in almost any transparent optical media at sufficiently high intensity; in air, for example, the orders range from $3$ to $8$ in the wavelength range of $248$–$800$ nm. Nonlinear Bessel beams play a prominent role in the filaments excited by Bessel beams, in which case multiphoton absorption produces ionization, but the weak plasma does not substantially alter the NBB structure \cite{POLESANA1,JUKNA,XIE}.

With Kerr nonlinearity and multiphoton absorption, the propagation of the complex envelope $A$ of a monochromatic light beam (or long enough pulse) $E=A\exp(-i\omega t +i k z)$ of angular frequency $\omega$ and of propagation constant $k=n\omega/c$, can be described by the nonlinear Schr\"odinger equation (NLSE)
\begin{equation}\label{NLSE}
\frac{\partial A}{\partial z} = \frac{i}{2 k} \Delta_\perp A + \frac{ik n_2}{n} \left| A \right|^2 A - \frac{\beta^{(K)}}{2} \left |A \right|^{2K-2} A \,,
\end{equation}
where $\Delta_\perp =\partial^2/\partial r^2 + (1/r)\partial/\partial r + (1/r^2)\partial^2/\partial \varphi^2$ is the transverse Laplacian in polar coordinates $(r,\varphi)$, $n$ and $n_2$ are the linear and nonlinear refractive indexes, $\beta^{(K)}>0$ is the nonlinear absorption coefficient, and $K$ the multiphoton absorption order in the medium at the selected frequency. Nonlinear Bessel beams feature, as their linear counterparts, a conical structure characterized by a half apex-angle, or cone angle $\theta$, and an associated shortening $\delta<0$ of the axial projection of the wave vector, given by $\delta = -k\theta^2/2$ in the paraxial approximation. Studying the dynamics of NBBs, the number of free parameters is minimized by introducing the normalized axial coordinate $\zeta =|\delta| z$, radial coordinate $\rho =\sqrt{k|\delta|} r$ and envelope $u=\sqrt{k|n_2|/n|\delta|}A$. The normalized NLSE in Eq. (\ref{NLSE}) then reads
\begin{equation}\label{NLSEN}
\frac{\partial u}{\partial \zeta} = \frac{i}{2}\Delta_\perp u \pm i|u|^2 u -\gamma |u|^{2K-2} u \,,
\end{equation}
where
\begin{equation}
\gamma = \frac{\beta^{(K)}}{2|\delta|}\left(\frac{n|\delta|}{k|n_2|}\right)^{K-1}
\end{equation}
specifies the strength of the nonlinear absorption relative to the Kerr nonlinearity, and the $\pm$ sign in Eq. (\ref{NLSEN}) is the sign of $n_2$.
In the figures we will also use normalized Cartesian coordinates $(\xi,\eta)=(\sqrt{k|\delta|}\,x,\sqrt{k|\delta|}\,y)$.

\begin{figure}[t!]
\centering
\vspace*{0.3cm}
\includegraphics[width=0.47\linewidth]{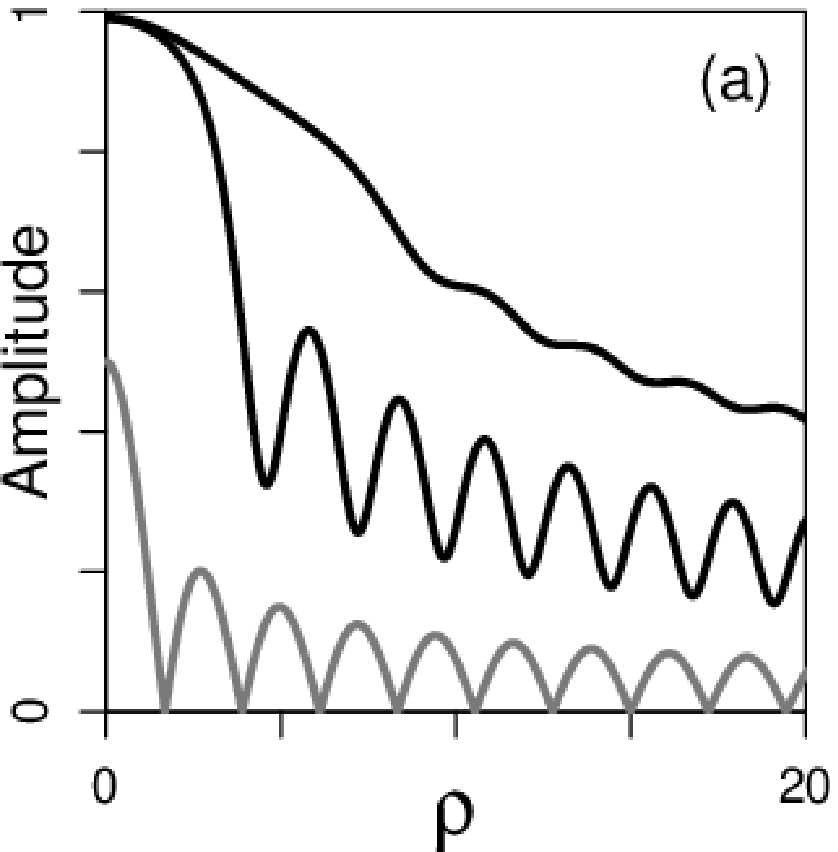}\hspace*{0.5cm}\includegraphics[width=0.47\linewidth]{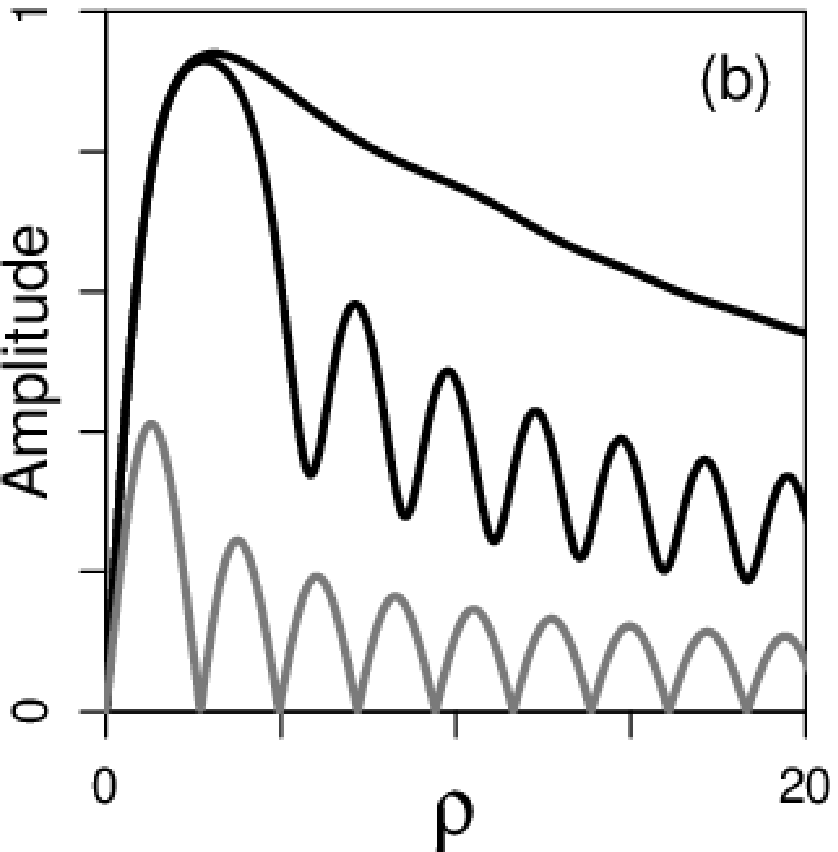}\\
\vspace*{0.3cm}
\includegraphics[width=0.47\linewidth]{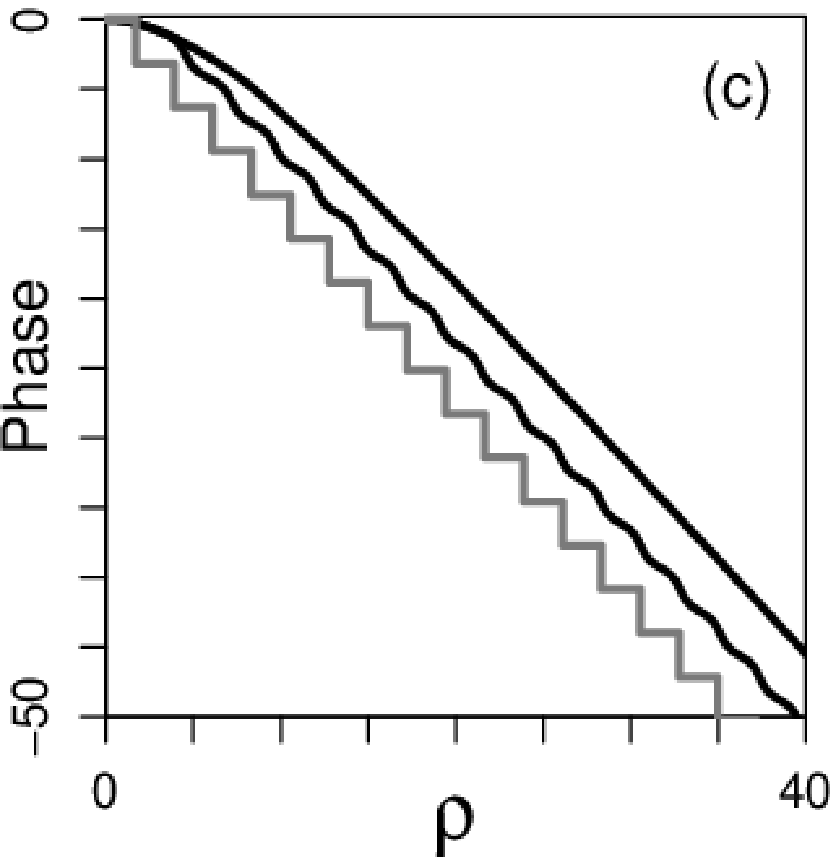}\hspace*{0.5cm}\includegraphics*[width=0.47\linewidth]{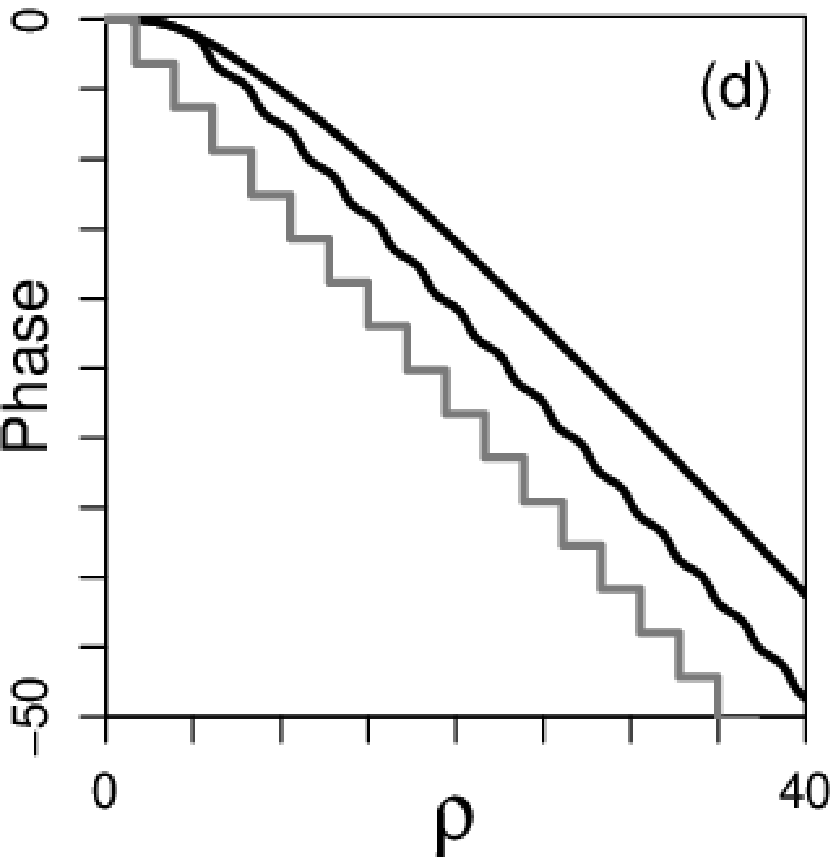}
\caption{\label{Fig1} Nonlinear Bessel beams in self-defocusing media with $\gamma=0.2$ and $K=4$. (a) Radial profiles of amplitude, $b(\rho)$, for $m=0$ with $C_0=0.991607, 0.99$ and $0.5$, the last one being indistinguishable from the linear Bessel beam with $m=0$ (gray curve). (b) Radial profiles of amplitude for $m=1$ with $C_1=0.821785, 0.821$ and $0.3535$, or linear Bessel beam with $m=1$. (c) and (d) Respective radial profiles of phase, $\phi(\rho)$.}
\end{figure}

\begin{figure}[t!]
\centering
\vspace*{0.15cm}
\includegraphics[width=0.47\linewidth]{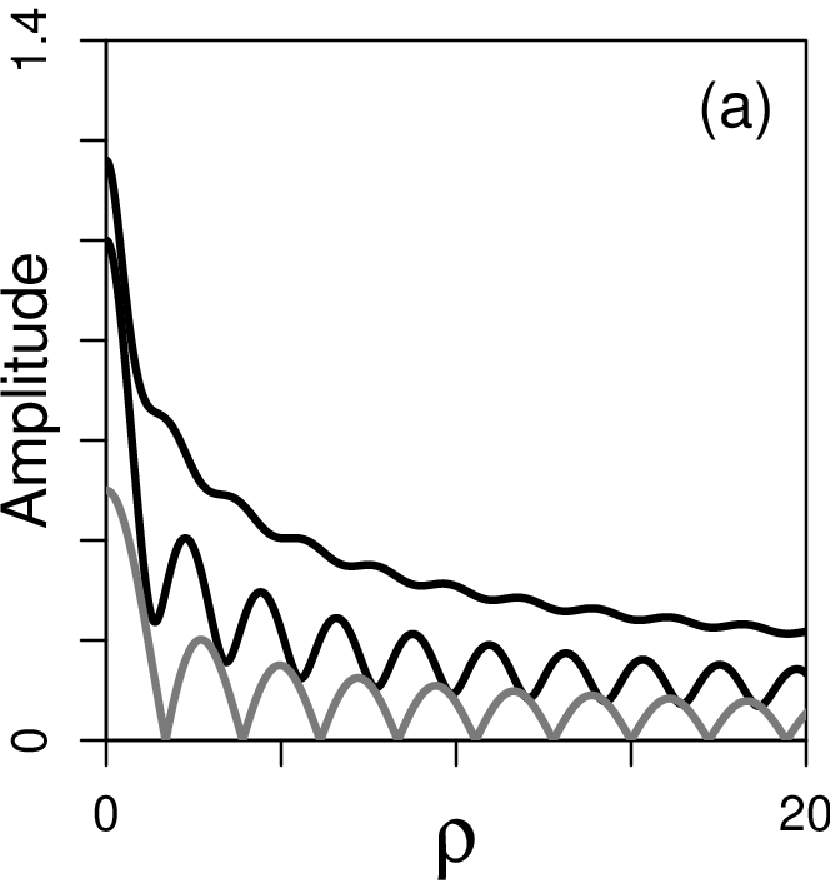}\hspace*{0.5cm}\includegraphics[width=0.47\linewidth]{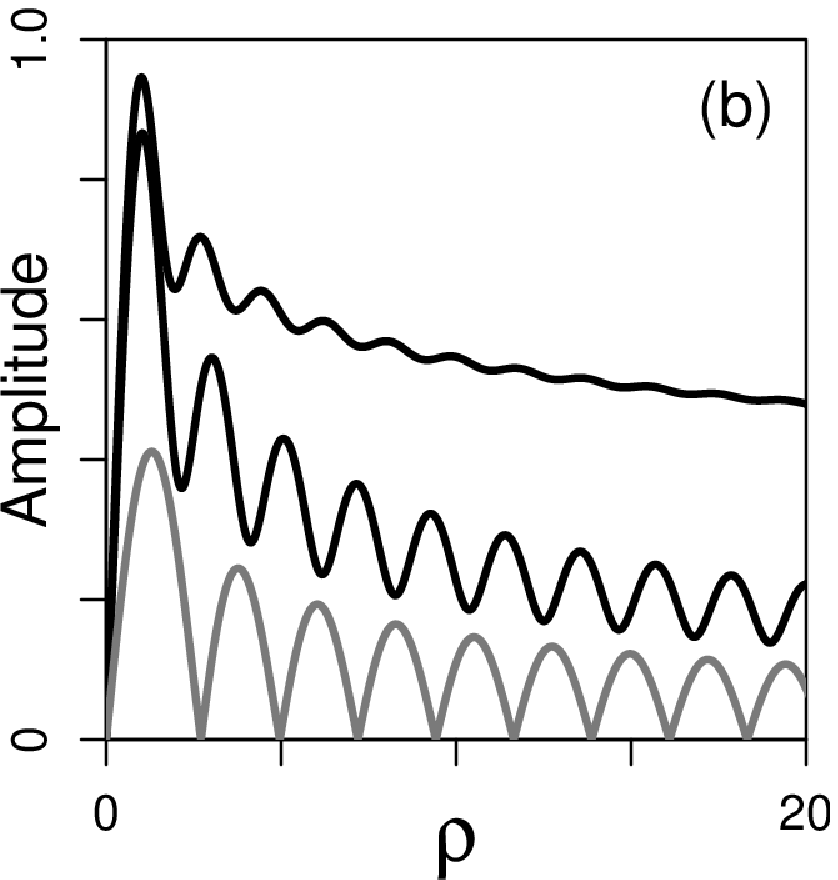}\\
\vspace*{0.3cm}
\includegraphics[width=0.47\linewidth]{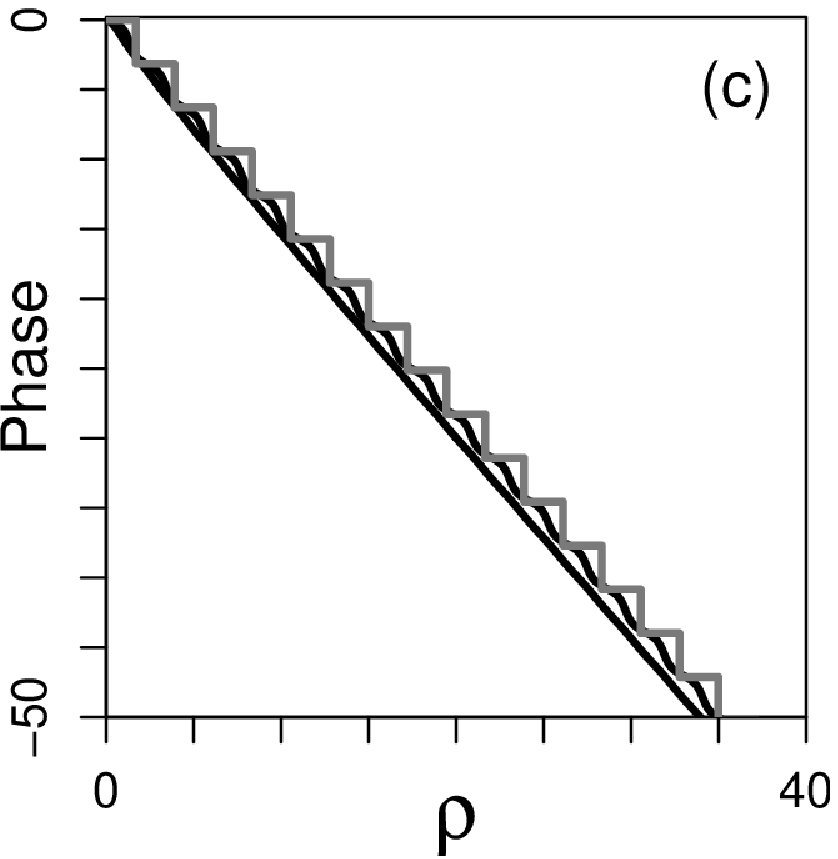}\hspace*{0.5cm}\includegraphics[width=0.47\linewidth]{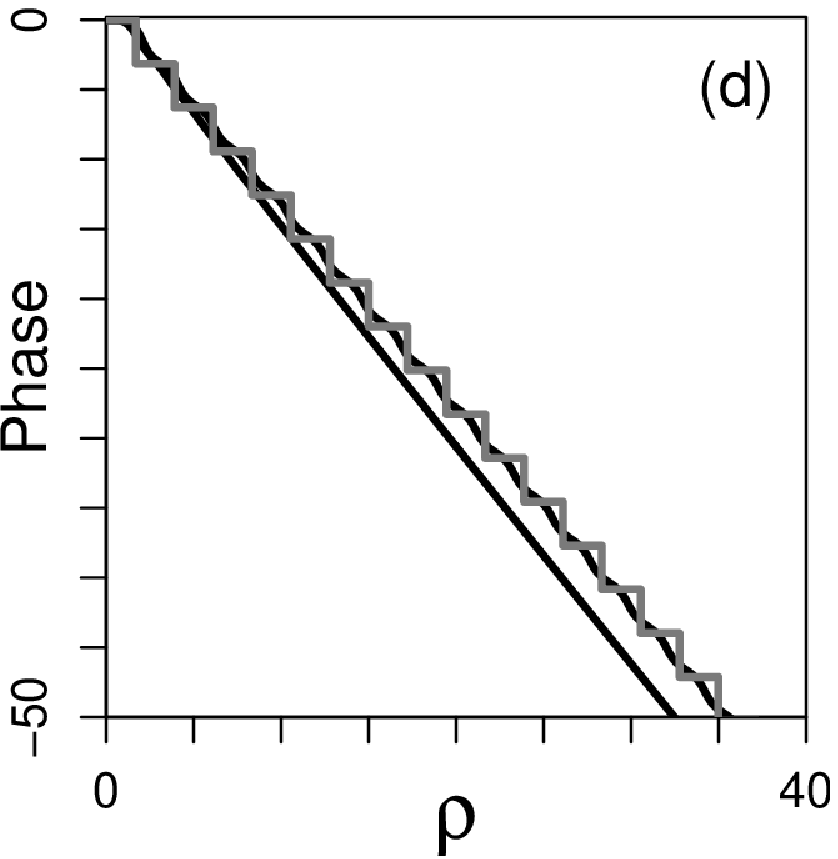}
\caption{\label{Fig2}  Nonlinear Bessel beams in self-focusing media with $\gamma=2$, $K=4$. (a) Radial profiles of amplitude, $b(\rho)$, for $m=0$ with $C_0=1.16, 1$ and $0.5$, the last one being indistinguishable from the linear Bessel beam with $m=0$ (gray curve). (b) Radial profiles of amplitude for $m=1$ with $C_1=1.39936, 1.25$ and $0.3535$, or linear Bessel beam with $m=1$. (c) and (d) Respective radial profiles of phase, $\phi(\rho)$.}
\end{figure}

 Nonlinear Bessel beams are propagation-invariant solutions to Eq. (\ref{NLSEN}) of the form $u_m= b(\rho)\exp[i\phi(\rho)] \exp(im\varphi)\exp(-i\zeta)$, where $\exp(-i\zeta)$ describes the axial wave vector shortening in these variables, the integer $m$ is the topological charge of the vortex at the beam center, and $b(\rho)>0$ and $\phi(\rho)$ are the real amplitude and phase radial profiles determined by the solutions of the ordinary differential equations
\begin{eqnarray}
\frac{d^2b}{d\rho^2} + \frac{1}{\rho}\frac{db}{d\rho}- \frac{m^2}{\rho^2}b - \left(\frac{d\phi}{d\rho}\right)^2 b + 2b \pm 2b^3&=&0\,, \label{AMP}\\
\frac{d^2\phi}{d\rho^2} + \frac{1}{\rho}\frac{d\phi}{d\rho} + 2 \frac{db}{d\rho}\frac{d\phi}{d\rho}\frac{1}{b}+ 2\gamma b^{2K-2}&=&0  \label{PHASE}\,,
\end{eqnarray}
with boundary conditions $b\sim C_{|m|}\rho^{|m|}$ as $\rho\rightarrow 0$ for a vortex of charge $m$, $b(\rho)\rightarrow 0$ as $\rho\rightarrow \infty$ as the condition of beam localization, $\phi(0)=0$ (an arbitrary phase) and $d\phi/d\rho|_{\rho=0}=0$. These solutions exist both in self-focusing and self-defocusing media up to a maximum value $C_{|m|,\rm max}$ of $C_{|m|}$ that depends on $m$, $K$ and $\gamma$. A few relevant examples are shown in Fig. \ref{Fig1} for self-defocusing media and in Fig. \ref{Fig2} for self-focusing media. For low enough $C_{|m|}$ the solutions approximate the linear Bessel beams. With increasing $C_{|m|}$ the amplitude profile becomes wider in self-defocusing media and narrower in self-focusing media, at the same time that the Bessel-like rings gradually lose their contrast, disappearing completely in the limit $C_{|m|}\rightarrow C_{|m|,\rm max}$ [Figs. \ref{Fig1}(a,b) and \ref{Fig2}(a,b)]. The feature that explains the vortex attractor property is the radial variation of the phase [Figs. \ref{Fig1}(c,d) and \ref{Fig2}(c,d)]. For linear Bessel beams, the variation is stairs-like with steps at each zero of the Bessel function. With nonlinear absorption the gradient of the phase radial profile is always negative, meaning that the phase front is permanently converging, with smooth steps at large radius. In the limit $C_{|m|} \rightarrow C_{|m|,\rm max}$ the steps disappear completely and the phase profile decreases linearly at large radius.

\begin{figure}[b!]
\centering
\includegraphics*[width=4.25cm]{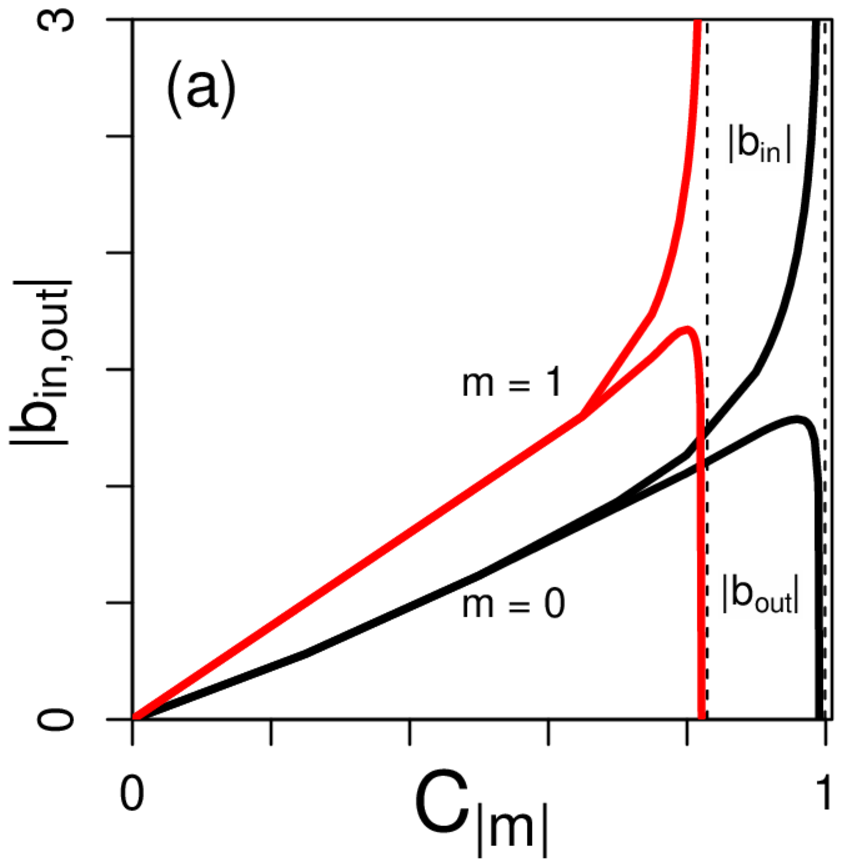}\includegraphics*[width=4.25cm]{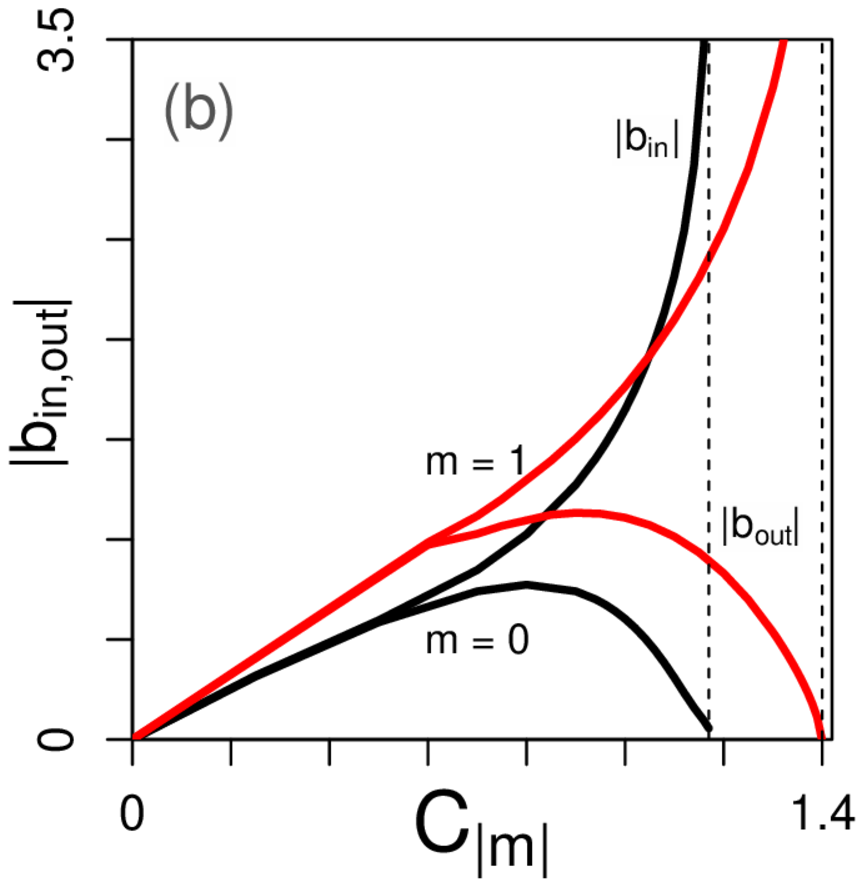}
\caption{\label{Fig3} Values of $|b_{\rm in}|$ and $|b_{\rm out}|$ of NBBs of charges $m=0$ and $|m|=1$ (a) in self-defocusing media with nonlinear absorption strength $\gamma=0.2$ and order $K=4$, and (b) in self-focusing media with $\gamma=2$ and order $K=4$, as functions of the parameter $C_{|m|}$ that characterizes the NBBs, up to the limit $C_{|m|,\rm max}$ of existence (vertical dashed lines).}
\end{figure}

Integration of Eq. (\ref{PHASE}) in $\rho$ yields
\begin{equation}\label{REFILLING}
F_\rho\equiv - 2 \pi\rho b^2\frac{d\phi}{d\rho} = 4\pi\gamma \int_0^{\rho} b^{2K} \rho d\rho \equiv N_\rho\,,
\end{equation}
a relation that explains the mechanism of stationarity in a medium with nonlinear absorption. The power-loss rate in any circle or radius $\rho$, $N_\rho$ equals to an equal inward-directed power-gain flowing through the circle circumference and coming from an intrinsic power reservoir in the Bessel-like tails. At large radius $\rho$, a NBB behaves in fact as the ``unbalanced" Bessel beam
\begin{equation}\label{ASYMP}
\resizebox{0.43\textwidth}{!}{$u_m \simeq \frac{1}{2}\left[b_{\rm out}H_m^{(1)}\left(\sqrt{2}\rho\right)+ b_{\rm in}H_m^{(2)}\left(\sqrt{2}\rho\right)\right]e^{im\varphi}e^{-i\zeta}\ $ }
\end{equation}
i. e., as a superposition of two H\"ankel beams carrying energy radially outwards and radially inwards of different amplitudes $b_{\rm out}$ and $b_{\rm in}$. In absence of nonlinear absorption $|b_{\rm out}|=|b_{\rm in}|$, and in absence of all nonlinearities $b_{\rm out}=b_{\rm in}$, in which case the right hand side of Eq. (\ref{ASYMP}) represents a linear Bessel beam. Use of Eq. (\ref{ASYMP}) in Eq. (\ref{REFILLING}) for $\rho\rightarrow\infty$ yields the relation
\begin{equation}
|b_{\rm in}|^2 - |b_{\rm out}|^2 = N_\infty \,,
\end{equation}
where $N_\infty = 4\pi\gamma \int_0^{\rho} b^{2K} \rho d\rho$ is the total power-loss rate in the transversal plane. Also, use of Eq. (\ref{ASYMP}) to evaluate asymptotically the intensity yields an expression of the form
\begin{equation}\label{HARMONIC}
\resizebox{0.43\textwidth}{!}{$2\pi\rho |u_m|^2 \simeq \frac{1}{\sqrt{2}}\left\{|b_{\rm in}|^2\!\!+\!\!|b_{\rm out}|^2\!\!+\!\!2|b_{\rm in}||b_{\rm out}| \cos[2(\sqrt{2}\rho + \Phi)]|\right\}\ $}
\end{equation}
where $\Phi$ is a constant phase, and which shows that $2\pi\rho |u_m|^2$ consists asymptotically at large radius of harmonic oscillations about a certain positive average value, the contrast of the oscillations being
\begin{equation}
C= \frac{2|b_{\rm in}||b_{\rm out}|}{|b_{\rm in}|^2 +|b_{\rm out}|^2}\, .
\end{equation}
Thus evaluation of $N_{\infty}$ and $C$ from the numerically obtained intensity profiles, allows us to obtain the inward and outward H\"ankel amplitudes as
\begin{equation}\label{BINBOUT}
|b_{\rm in,out}|^2 = \frac{N_\infty}{2}\left[\sqrt{\frac{1}{1-C^2}}\pm 1\right]\,.
\end{equation}
As examples, Fig. \ref{Fig3} shows the values of $|b_{\rm in}|$ and $|b_{\rm out}|$ as functions of $C_{|m|}$ for different topological charges in self-defocusing media with $\gamma=0.2$ and $K=4$ and in self-focusing media with $\gamma=2$ and $K=4$.

While NBBs in transparent media ($\gamma=0$) suffer from instability in both self-focusing \cite{PORRAS2} and self-defocusing media \cite{PORRAS3}, a sufficient amount of nonlinear absorption stabilizes them against small perturbations. In self-focusing media the threshold for stability is of the order of $\gamma\sim 1$ \cite{PORRAS2}, the exact value depending on the particular value of $C_{|m|}$, $m$ and $K$. In self-defocusing media this threshold is found to be lower and given by $\gamma \sim 0.1$.

Regarding the practical generation of NBBs, they are excited by linear Bessel beams introduced in the nonlinear medium \cite{JUKNA,RUIZ,XIE,POLESANA1,POLESANA2,POLESANA3,POLESANA4}. The linear Bessel beam $u= b_B J_m\left(\sqrt{2}\rho\right)\exp(im\varphi)$ reshapes into the NBB characterized by the same cone angle, topological charge $m$ and  with inward H\"ankel beam amplitude $|b_{\rm in}|$ equal to $b_B$ \cite{RUIZ,PORRAS4}. Since for the linear Bessel beam $|b_{\rm in}|=|b_{\rm out}|=b_B$, the NBB that tends to be formed is that preserving the amplitude of the inward H\"ankel beam component. Thus the specific NBB (the value of $C_{|m|}$) can be determined from Fig. \ref{Fig3} as that whose value of $|b_{\rm in}|$ is $b_B$.

In actual experiments, finite-power Bessel beams generated by axicons or spatial light modulators are used \cite{XIE,POLESANA1}. Starting with a Gaussian beam $u=b_G \exp{(-\rho^2/\rho_0^2)}$ [or $A=B_G \exp(-r^2/r_0^2)$ with physical variables] and a small or punctual vortex $\exp(im\varphi)$ at its center, and after passage through the axicon imprinting the conical phase $\exp(-i\sqrt{2}\rho)$ [or $\exp(-ikr\theta)$ with physical variables], a finite-power or apodized version of the linear Bessel $u= b_B J_m(\sqrt{2}\rho)\exp(im\varphi)\exp(-i\zeta)$ [or $A=B_BJ_m(k\theta r)\exp(im\varphi)\exp(i\delta z)$] of amplitude $b_B^2 = b_G^2 \pi \rho_0\sqrt{2/e}$ [or $B_B^2 = B_G^2 \pi k\theta r_0/\sqrt{e}$] would be formed in free space about the center $\rho_0/2\sqrt{2}$ [$r_0/2\theta$] of the Bessel zone of length $\zeta_B=\rho_0/\sqrt{2}$ [$z_B=r_0/\theta$] after the axicon \cite{POLESANA1}. In a nonlinear medium placed immediately after the axicon, instead, the NBB with $|b_{\rm in}|=b_B$ [or $|B_{\rm in}|\equiv |b_{\rm in}|\sqrt{n|\delta|/k|n_2|}=B_B$] is formed in the Bessel zone, i. e., the NBB  preserving again the inward H\"ankel beam amplitude \cite{PORRAS2,PORRAS4}.

\section{Nonlinear Bessel beams as optical vortex attractors and their applications}

In this section we consider the dynamics of vortices embedded in a NBB, in either self-defocusing or self-focusing media. Since a NBB may contain a vortex at its center, we must make the distinction between the ``intrinsic" vortex of the NBB and the ``foreign" vortices.

First, the existence of stable NBBs with an intrinsic vortex of rather arbitrary charge in its center means that the vortex will remain there when the NBB is subjected to small perturbations \cite{PORRAS2}. This is in contrast to a vortex nested in the center of a Gaussian-like or super-Gaussian-like beam, where any perturbation and diffraction make the vortex to move radially outwards, decay and broaden \cite{ROZAS}. Also in sharp contrast to the complex vortex dynamics of foreign vortices in Gaussian or super-Gaussian beams \cite{ROZAS,KIVSHAR,KIVSHAR2}, this dynamics in NBBs is quite simple. Provided that the strength of nonlinear absorption $\gamma$ makes the $n$-vortex NBB robust enough against the disturbance of $N$ vortices of total charge $s$ placed arbitrarily, the $N$ vortices end in the beam center, combining with the intrinsic vortex of the $n$-vortex NBB, at the same time that the NBB transforms into a $m$-vortex NBB preserving the topological charge, i. e., with $m=n+s$. The specific $m$-vortex NBB is that preserving also the value of $|b_{\rm in}|$. Additional weak perturbations in the NBB do not alter this vortex attraction property.

\begin{figure}[th!]
\centering
\includegraphics*[width=3.5cm]{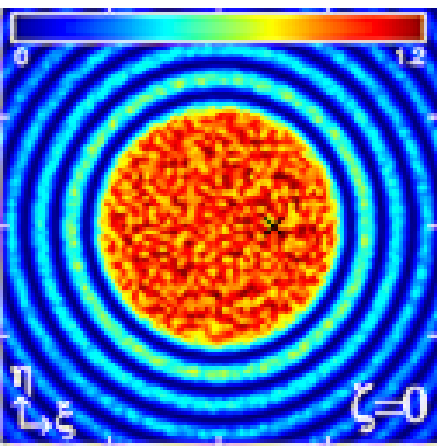}\includegraphics*[width=3.5cm]{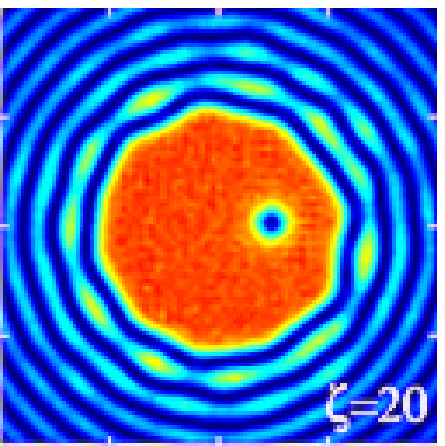}
\includegraphics*[width=3.5cm]{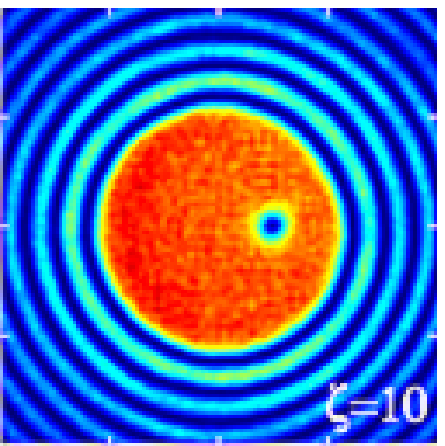}\includegraphics*[width=3.5cm]{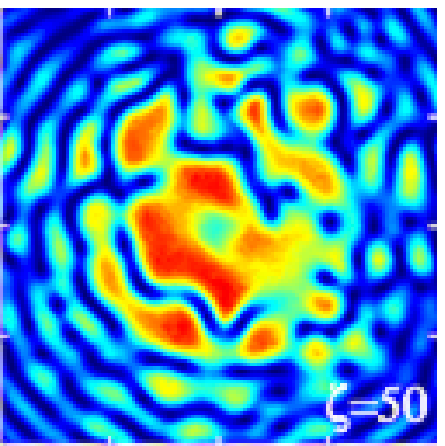}
\caption{\label{Fig4} {\it Instability of the NBB background without nonlinear absorption.} Transversal intensity profiles at the indicated distances of the fundamental NBB ($n=0$) with $C_0=0.99999999$ in a self-defocusing transparent medium ($\gamma=0$) with a single-charged, punctual vortex (black cross) placed at $(\xi,\eta)=(5,0)$ at $\zeta=0$. Random noise of $10\%$ the peak amplitude is introduced to recreate more realistic conditions. The vortex remains at rest at its original location but instability of the NBB causes the destruction of the NBB-vortex system. Distance between ticks is $10$.}
\end{figure}

If for simplicity we ignore the mutual influence of the foreign vortices, the attraction property of each vortex can be understood from the well-known laws of the vortex dynamics in a background field \cite{KIVSHAR,KIVSHAR2}. A vortex placed at a distance $\rho$ in a background field of amplitude $b$ and phase $\phi$ acquires a radial velocity equal to the radial gradient of the phase, i. e., $v_\rho(\rho) \equiv d\rho/d\zeta = \partial \phi/\partial\rho$ (note the opposite sign in Ref. \cite{KIVSHAR,KIVSHAR2} due to the opposite sign of the complex representation of harmonic fields). Since the radial gradient of NBBs is negative at any distances, the vortex at any position acquires and inward-directed radial velocity. At large radius Eq. (\ref{REFILLING}) yields $\partial \phi/\partial\rho\simeq - N_\infty/2\pi\rho b^2$, and from Eqs. (\ref{HARMONIC}) and (\ref{BINBOUT}) we obtain the radial velocity
\begin{equation}\label{VEL}
v_\rho(\rho) = \frac{\partial \phi}{\partial\rho} =  -\sqrt{2} \frac{\sqrt{1-C^2}}{1+C\cos[2(\sqrt{2}\rho + \Phi)]}\,.
\end{equation}
We can also estimate the distance that the vortex placed at $\zeta=0$ at the radial distance $\rho_V$ takes to reach the beam central part of the NBB  ($\rho\simeq 0$). Integrating Eq. (\ref{VEL}), we obtain $\left.\rho + (C/2\sqrt{2})\sin\left[2\left(\sqrt{2}\rho +\Phi\right)\right]\right|^0_{\rho_V} = -\sqrt{2}\sqrt{1-C^2}\zeta$, and neglecting the term with the sinusoidal term in comparison to the large $\rho_V$,
\begin{equation}\label{DIST}
\zeta\simeq \frac{\rho_V}{\sqrt{2(1-C^2)}}\, .
\end{equation}
This distance is minimum, $\zeta \simeq \rho_v/\sqrt{2}$, with $C$ approaching zero, i. e., with NBBs with $C_{|n|}\rightarrow C_{|n|,\rm max}$, or higher possible intensity. With physical variables, this minimum distance is $z=r_v/\theta$, i. e., the vortex follows in this case the conical flow of power in conical beams of the cone angle $\theta$. Of course, once the vortex approaches the nonlinear center, where the asymptotic expressions are not valid, it continues to approach the center since the phase gradient is also negative in this region, but the velocity can only be evaluated  numerically.

Thus nonlinear absorption plays a two-fold role: It stabilizes the NBB, and bends the phase fronts for the vortices to be pushed towards the beam center. In absence of nonlinear absorption, Eq. (\ref{DIST}) with $C=1$ yields $\zeta\rightarrow \infty$, in agreement with the fact that nested vortices do not appreciably acquire a radial velocity in linear Bessel beams \cite{VBESSEL}. Also, as recently shown in \cite{PORRAS3}, and as illustrated in Fig. \ref{Fig4}, a foreign vortex in a NBB of the fully transparent, self-defocusing medium remains at rest, surviving for longer distances compared to Gaussian or super-Gaussian backgrounds, but instability of the NBB leads to the destruction of the NBB-vortex system. In a transparent self-focusing medium the situation is more dramatic, since the self-focusing instability of NBBs is severe. The presence of noise or introducing any foreign vortex triggers the instability that quickly destroys the system.

As we envisage quite different applications of the vortex attractive property in nonlinearly absorbing media, we demonstrate it numerically with the different values of $n$ of the initial NBB and of the $N$ foreign vortices of total charge $s$ that are relevant to these applications. To ensure stability of the NBBs we take $\gamma$ well-above $0.1$ and $1$ in the respective cases of self-defocusing and self-focusing media. Also, in order to speed up the vortex motion towards the NBB center we will use NBBs with $C_{|n|}$ close to the limit $C_{|n|,\rm max}$ of existence, i. e., of low radial contrast. In all simulations we additionally introduce random noise in the complex amplitude to recreate realistic conditions.
In the first examples we consider NBBs with their intrinsic infinite-power reservoir for simplicity. In order to show that the attractive vortex property and their applications hold with physically realizable, finite-power (apodized) NBBs, in the last example we consider the massive annihilation of the vortices carried by a speckled Gaussian beam that is transformed by an axicon and by propagation in the nonlinear medium into a physically realizable NBB.

\subsection{Vortex trapping with mode conversion}

\begin{figure}[th!]
\centering
\includegraphics*[width=3.5cm]{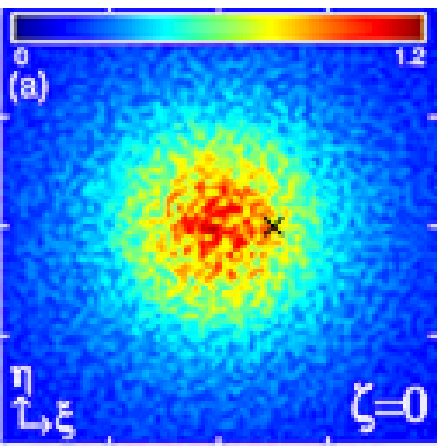}\includegraphics*[width=3.5cm]{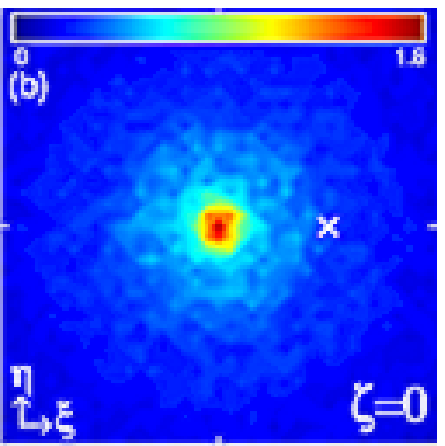}\\
\includegraphics*[width=3.5cm]{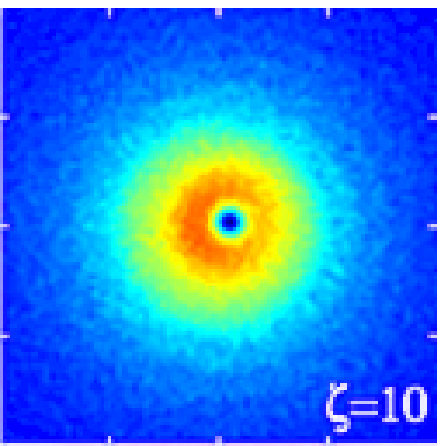}\includegraphics*[width=3.5cm]{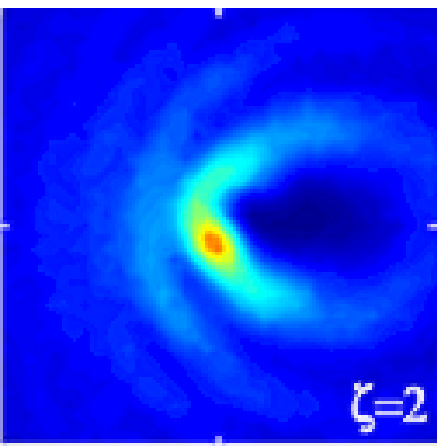}\\
\includegraphics*[width=3.5cm]{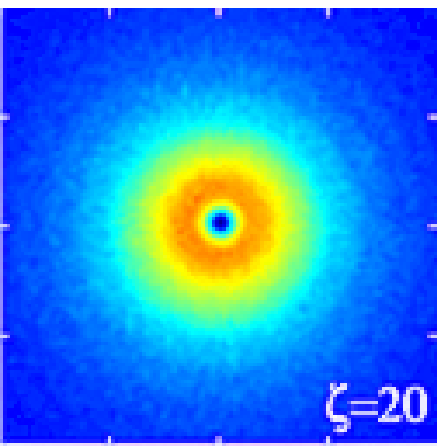}\includegraphics*[width=3.5cm]{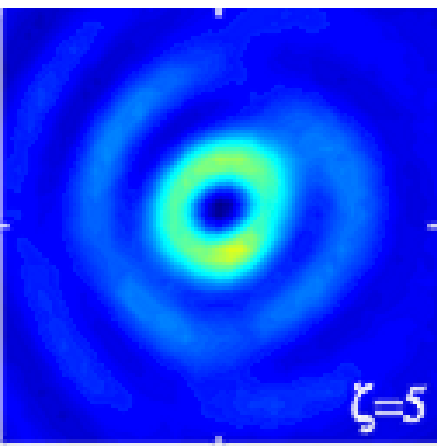}\\
\includegraphics*[width=3.5cm]{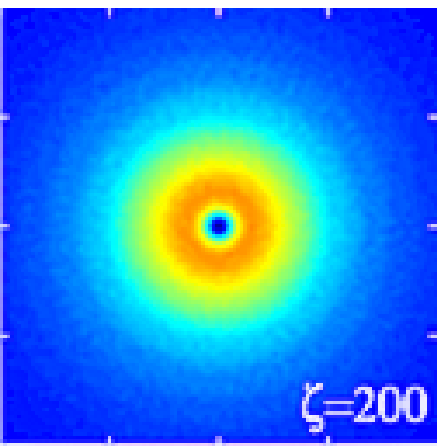}\includegraphics*[width=3.5cm]{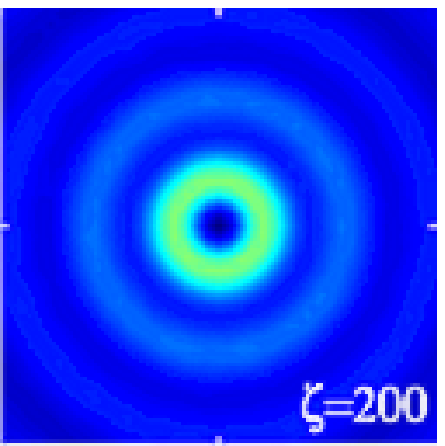}
\caption{\label{Fig5} {\it Vortex trapping and stability with nonlinear absorption:} Transversal intensity profiles at the indicated distances of (a) NBB with $n=0$ and $C_0=0.991607$ ($|b_{\rm in}|=6.82671$) in a self-defocusing, nonlinearly absorbing medium with $\gamma=0.2$ and $K=4$, and of (b) NBB with $n=0$ and $C_0=1.16$ ($|b_{\rm in}|=3.469$) in a self-focusing, nonlinearly absorbing medium with $\gamma=2$ and $K=4$. In the two cases a punctual single-charged vortex (crosses) is placed at $(\xi,\eta)=(5,0)$ at $\zeta=0$, and random noise of maximum value $10\%$ the peak amplitude is introduced. Distance between ticks is $10$. In the two cases, the vortices are trapped in the NBB center, where they propagate stably.}
\end{figure}

\begin{figure}[th!]
\centering
\includegraphics*[width=4cm]{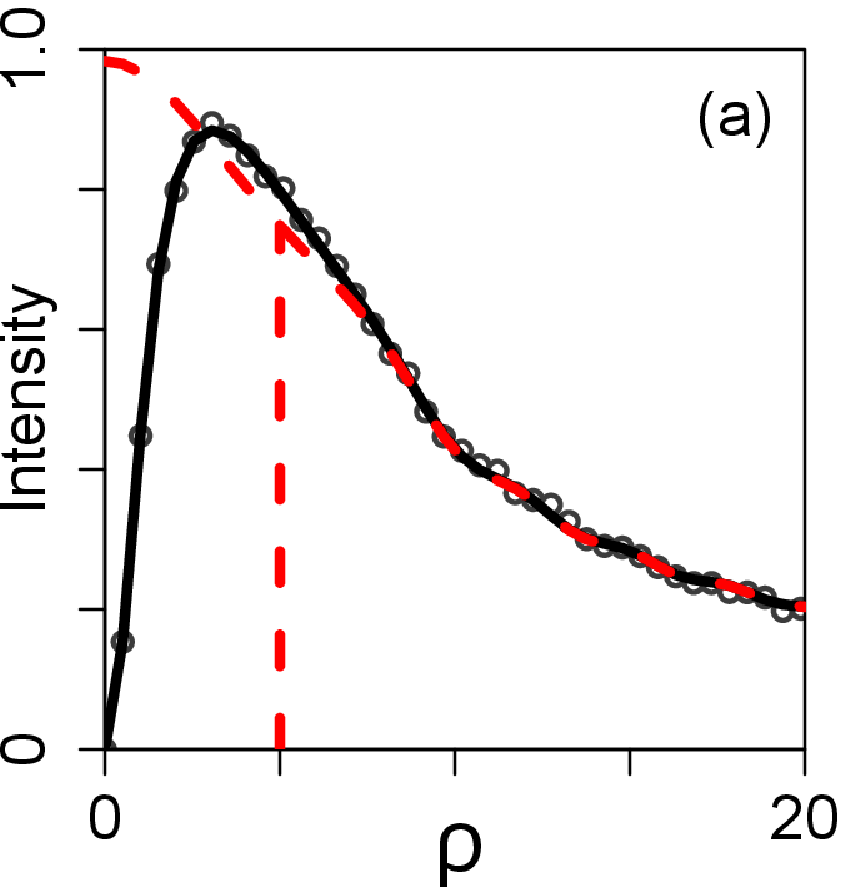}\hspace*{0.5cm}\includegraphics*[width=4cm]{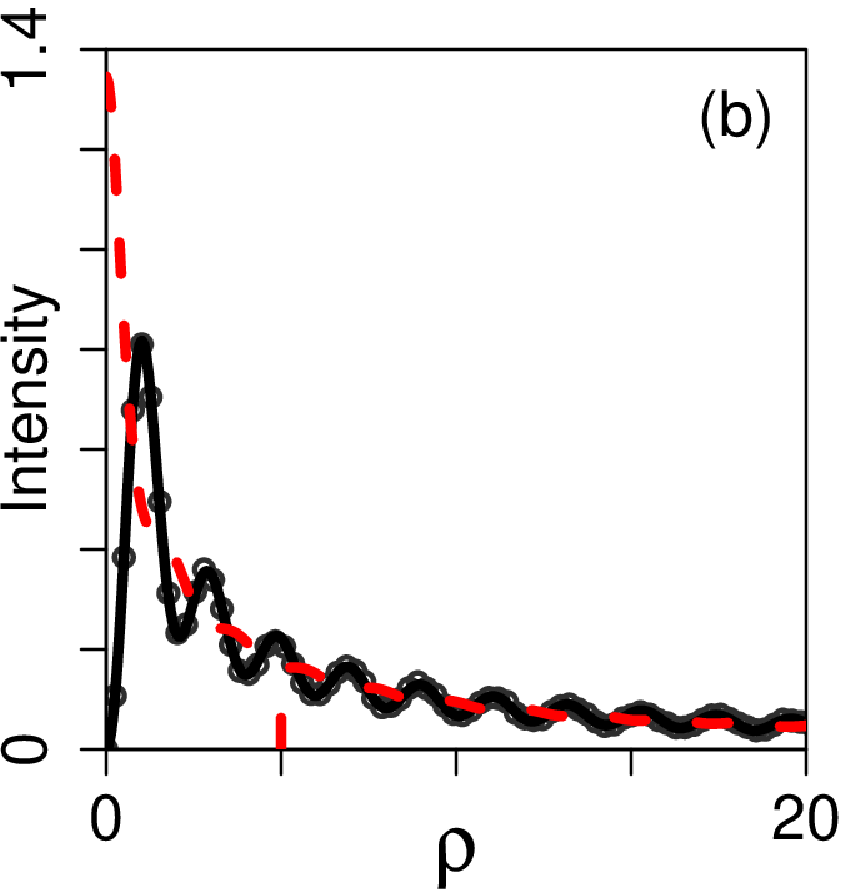}
\caption{\label{Fig6}{\it Mode conversion:} Red dashed curves: Radial intensity profiles of input NBBs with (a) $n=0$ and $C_0=0.991607$ ($|b_{\rm in}|=6.82671$) in a self-defocusing, nonlinearly absorbing medium with $\gamma=0.2$ and $K=4$ in self-defocusing medium, (b) $n=0$ and $C_0=1.16$ ($|b_{\rm in}|=3.469$) in a self-focusing, nonlinearly absorbing medium with $\gamma=2$ and $K=4$. The position $(\xi,\eta)=(5,0)$ of the punctual vortex is indicated by vertical red dashed lines. Open circles: corresponding final, stable states at sufficiently long propagation distance. Black curves: Radial intensity profiles of NBBs with (a) $m=1$ and $C_1=0.82178$ such that $|b_{\rm in}|=6.82671$ is conserved in the self-defocusing medium and with (b) $m=1$ and $C_1= 1.3197$ such that $|b_{\rm in}|=3.469$ is also conserved in the self-focusing medium. Fitting of the final states to these NBBs demonstrates mode conversion that preserves the topological charge and $|b_{\rm in}|$.}
 \end{figure}

Figures \ref{Fig5}(a) and \ref{Fig5}(b) illustrate the trapping of an off-axis vortex with $s=1$ by NBBs in self-defocusing and self-focusing media with nonlinear absorption. The NBB backgrounds resist the large deformation produced by the penetration of the vortex in the NBB center, and propagate stably once the vortex has been placed at the center. The nature of the final stable state is clear from Figs. \ref{Fig6}(a) and \ref{Fig6}(b). For respective self-defocusing and self-focusing media, the dashed curves represent the radial intensity profiles of the input NBBs with $n=0$ and the punctual vortex placed at a certain distance from the center (vertical line), and the open circles represent the radial intensity profiles of the final stable states. The later are indistinguishable from the black curves, representing the radial intensity profiles of NBBs with $m=1$ with the same strengths of nonlinear absorption $\gamma$ and with $C_1$ such that $|b_{\rm in}|$ is the same as for the input NBBs with $n=0$, as obtained from Fig. \ref{Fig3}. In other words, the NBBs with $n=0$ and with the nested vortices have experienced a conversion to NBBs with $m=1$ preserving the topological charge and the amplitudes of the inward H\"ankel beam components. Similar vortex trapping and mode conversion are numerically observed with a multiply-charged vortex ($N=1$, $|s|>1$) placed anywhere in NBBs of any order $n$: The final state is the NBB of order $m=n+s$ with the same $|b_{\rm in}|$ as the NBB of the original charge $n$.

Thus, in order to generate a high-order NBB for applications such as tubular filamentation in \cite{XIE}, precise alignment of the phase masks imprinting the phase dislocations is not necessary. Also, once the final $m$-vortex NBB exits from the nonlinear medium and propagates in free space, it will transform into the $m$-vortex lineal Bessel beam that preserves, as above, the amplitude of the inward H\"ankel component, i. e., into $|b_{\rm in}| J_m(\sqrt{2}\rho) e^{im\varphi}e^{-i\zeta}$. Thus, starting with the fundamental NBB (generated, for instance, by an axicon), nesting randomly a vortex of charge $s$, and making the whole to propagate a certain distance in a nonlinearly absorbing medium, would be an alternative method to the one developed in \cite{ZUKAUSKAS}, with a monolithic, careful aligned axicon and spiral phase, for the generation of high-order Bessel beams.

\subsection{Vortex combination}

\begin{figure}[th!]
\centering
\includegraphics*[width=3.5cm]{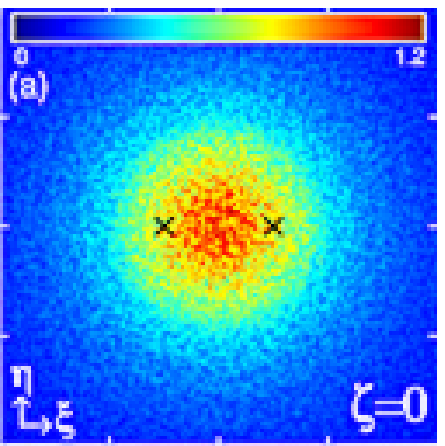}\includegraphics*[width=3.5cm]{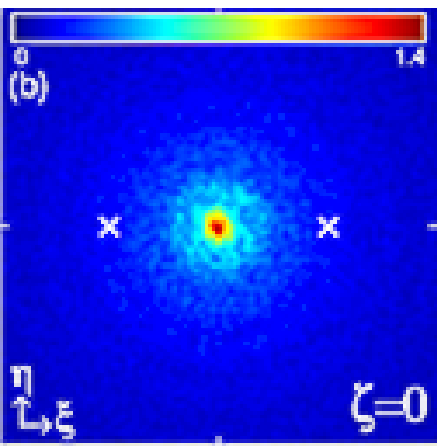}\\
\includegraphics*[width=3.5cm]{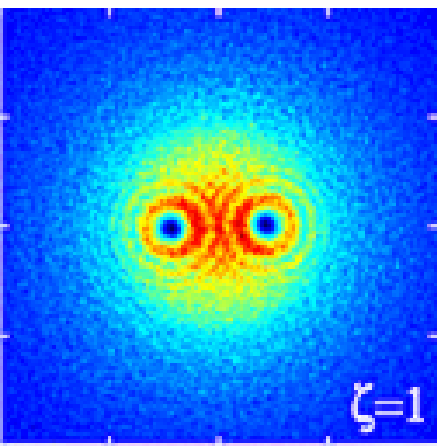}\includegraphics*[width=3.5cm]{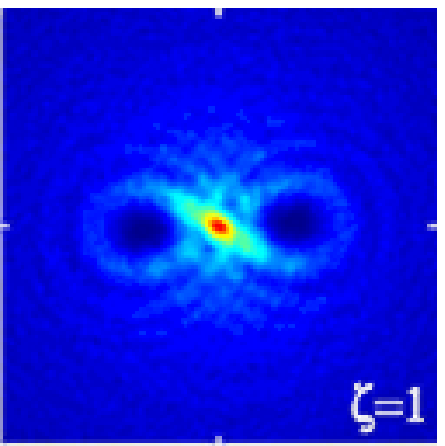}\\
\includegraphics*[width=3.5cm]{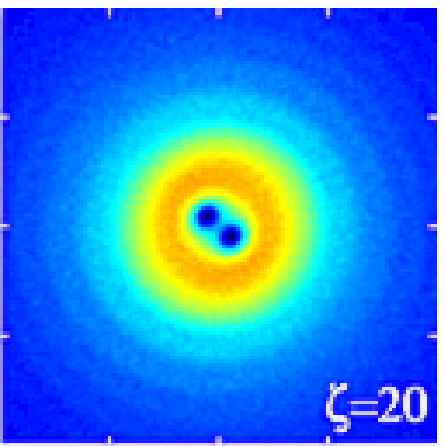}\includegraphics*[width=3.5cm]{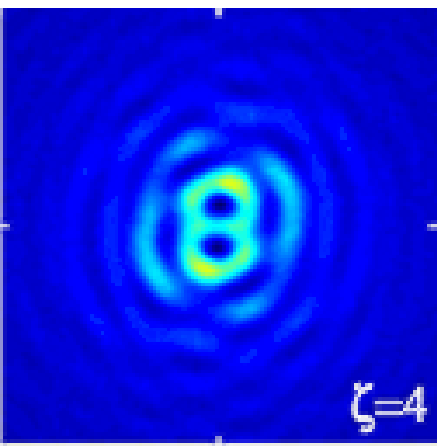}\\
\includegraphics*[width=3.5cm]{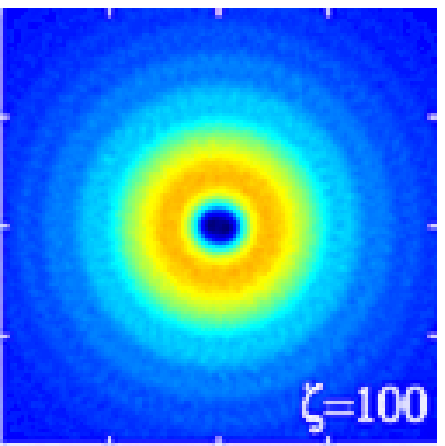}\includegraphics*[width=3.5cm]{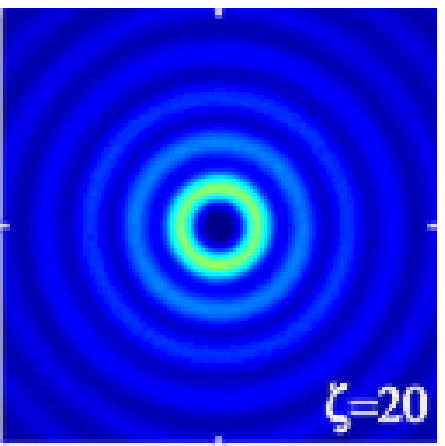}
\caption{\label{Fig7} {\it Vortex combination:} Transversal intensity profiles at the indicated propagation distances of (a) NBB with $n=0$ and $C_0=0.991607$ in a self-defocusing, nonlinearly absorbing medium with $\gamma=0.2$ and $K=4$, and (b) NBB with $n=0$ and $C_1=1.16$ in a self-focusing, nonlinearly absorbing medium with $\gamma=2$ and $K=4$. In the two cases two equal point vortices (black/white crosses) with $s_1=s_2=1$ are placed at $(\xi,\eta)=(5,0)$ and at $(\xi,\eta)=(-5,0)$ at $\zeta=0$, and random noise of maximum value $10\%$ the peak amplitude is introduced. Distance between ticks is $10$. In the two cases, the vortices combine in the NBB centers, forming NBBs with $m=2$ in the corresponding media that preserve the amplitudes of the inward H\"ankel beam components.}
\end{figure}

Two equal vortex solitons separated a certain distance in an infinite plane wave background in a self-defocusing medium rotate indefinitely \cite{KIVSHAR}, while vortices in a Gaussian or super-Gaussian background spiral out, broaden and decay \cite{ROZAS,ROZAS2,ROZAS3}. The opposite is true when nested in a NBBs in media with nonlinear absorption. As seen in Fig. \ref{Fig7}(a) and (b) for the respective cases of self-defocusing and self-focusing media, the two equal vortices ($N=2$) of charges $s_1=s_2=1$ ($s=2$) in the NBB with $n=0$ combine in the center, forming the NBB with $m=2$ that also preserves the amplitude of the inward H\"nkel beam component. Similar dynamics is observed with other couples of foreign vortices of equal or unequal charges in NBBs of different orders $n$.

\begin{figure}[th!]
\centering
\includegraphics*[width=3.5cm]{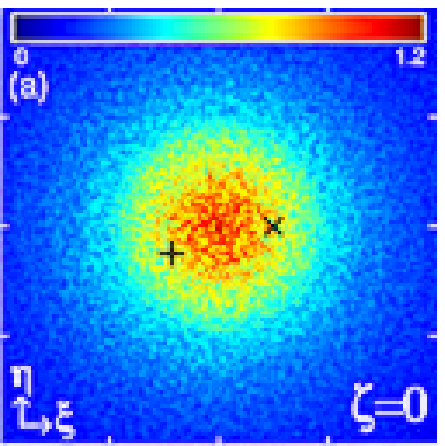}\includegraphics*[width=3.5cm]{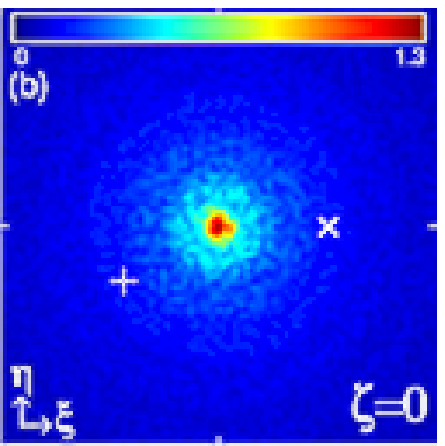}\\
\includegraphics*[width=3.5cm]{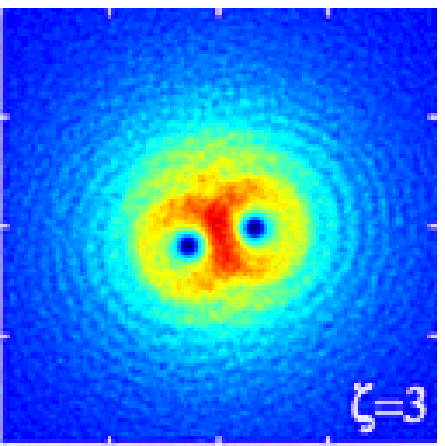}\includegraphics*[width=3.5cm]{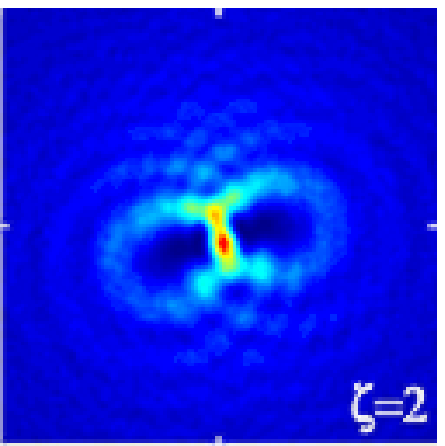}\\
\includegraphics*[width=3.5cm]{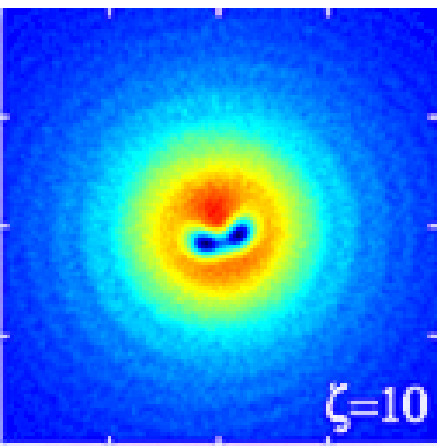}\includegraphics*[width=3.5cm]{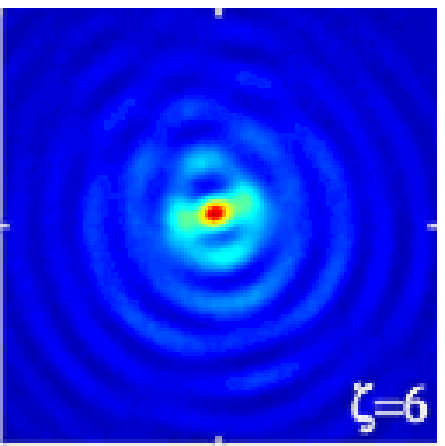}\\
\includegraphics*[width=3.5cm]{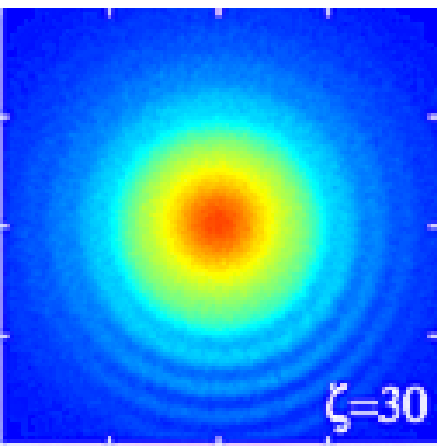}\includegraphics*[width=3.5cm]{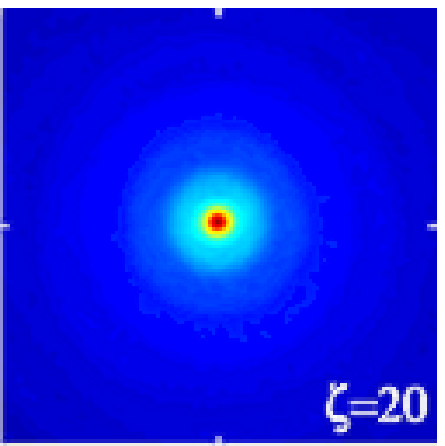}
\caption{\label{Fig8} {\it Vortex dipole annihilation:} Transversal intensity profiles at the indicated propagation distances of (a) NBB with $n=0$ and $C_0=0.991607$ in a self-defocusing, nonlinearly absorbing medium with $\gamma=0.2$ and $K=4$, and (b) NBB with $n=0$ and $C_1=1.16$ in a self-focusing, nonlinearly absorbing medium with $\gamma=2$ and $K=4$. In the two cases two vortices of opposite charges $s_1=-s_2=1$ are nested at $\zeta=0$ at the indicated transversal positions (black/white crosses), and random noise of maximum value $10\%$ the peak amplitude is introduced. Distance between ticks is $10$. In the two cases, the vortex dipole annihilate, and the final state is equal to the initial state without the dipole and without noise.}
\end{figure}

\subsection{Vortex dipole annihilation}

Of particular relevance is a vortex dipole, or two vortices of opposite charges. The vortices of the dipole move along parallel trajectories when placed in a infinite plane wave background \cite{KIVSHAR}. In a Gaussian beam the dipole dynamics is extremely complex; they may annihilate or not depending on the particular positions, annihilation being desirable for most of applications \cite{RUX}. In the NBB with $n=0$ the two vortices of the dipole always merge, irrespective of their initial positions, in the NBB center and annihilate, the final state being the initial NBB without the nested vortices, that have been removed from the background beam with this procedure. This is illustrated in Fig. \ref{Fig8} (a) and (b) for the respective cases of self-defocusing and self-focusing and for a vortex dipole ($N=2$) with $s_1=-s_2=1$ ($s=0$) in the fundamental NBB.

\subsection{Light beam cleaning}

\begin{figure}[th!]
\centering
\includegraphics[width=0.4\linewidth]{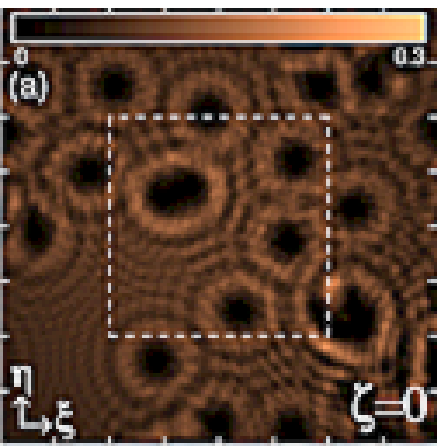}\includegraphics[width=0.4\linewidth]{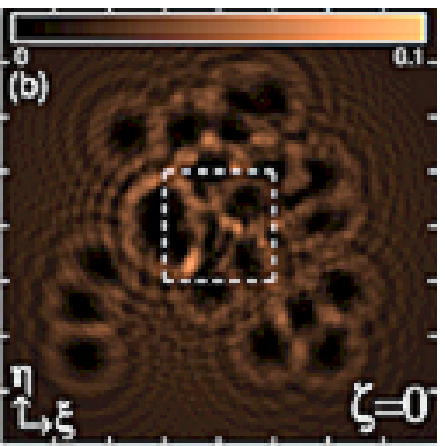}
\includegraphics[width=0.4\linewidth]{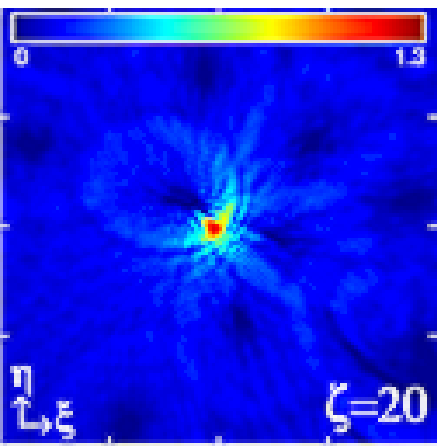}\includegraphics[width=0.4\linewidth]{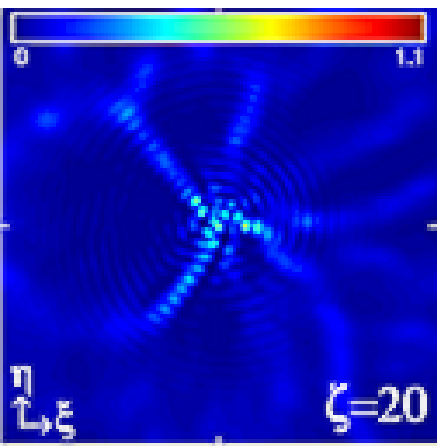}
\includegraphics[width=0.4\linewidth]{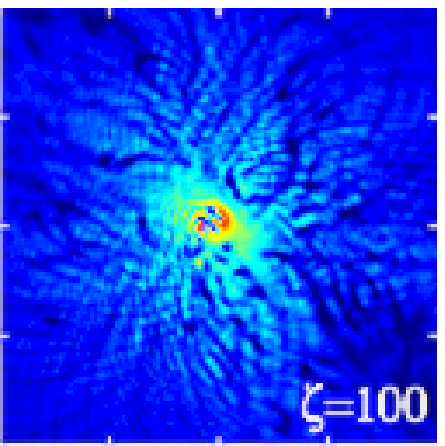}\includegraphics[width=0.4\linewidth]{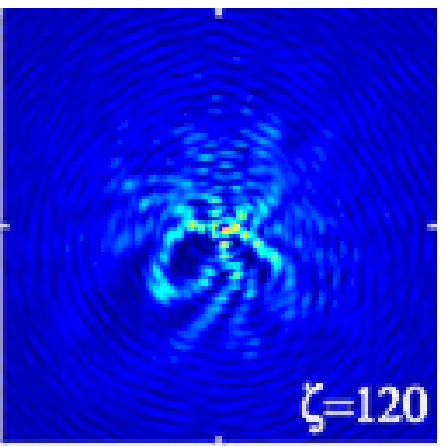}
\includegraphics[width=0.4\linewidth]{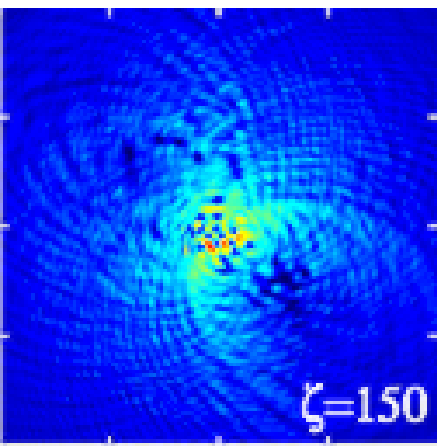}\includegraphics[width=0.4\linewidth]{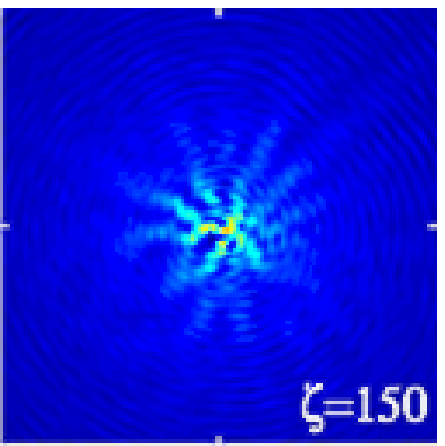}
\includegraphics[width=0.4\linewidth]{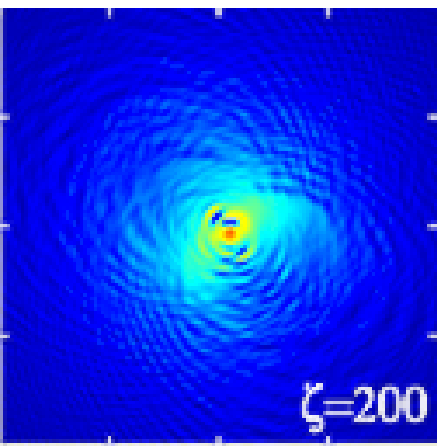}\includegraphics[width=0.4\linewidth]{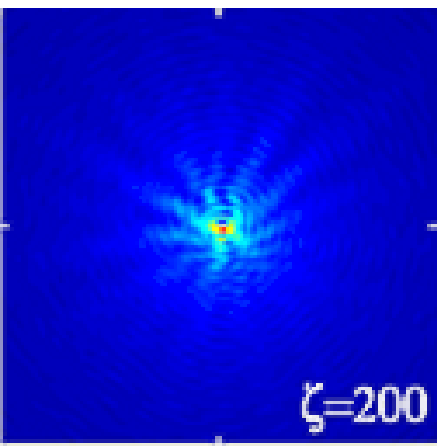}
\includegraphics[width=0.4\linewidth]{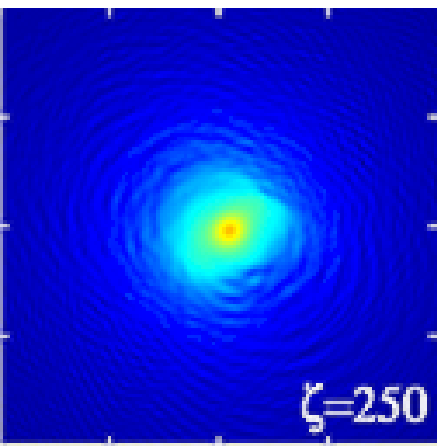}\includegraphics[width=0.4\linewidth]{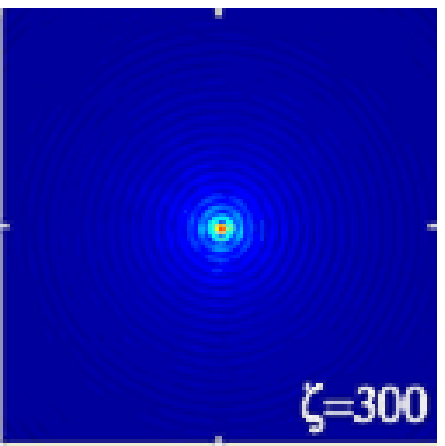}
\caption{\label{Fig9} {\it Beam cleaning.} Top panels: Intensity of central portions of Gaussian beams of width $\rho_0=400$, and (a) amplitude $b_G=0.3$ with $N=50$ randomly placed vortices of charges $+1$ and $-1$, (b) amplitude $b_G=0.15$ with $N=24$ randomly placed vortices, illuminating an axicon. Lower panels: Intensity at the indicated distances in (a) self-defocusing medium with $K=4$, $\gamma=0.2$, (b) self-focusing medium with $K=4$, $\gamma=2$, placed after the axicon. All vortices are washed out about the end of the Bessel zone of length $\rho_0/\sqrt{2}=281$. Distance between ticks is $40$.}
\end{figure}

\begin{figure}[th!]
\centering
\includegraphics[width=0.4\linewidth]{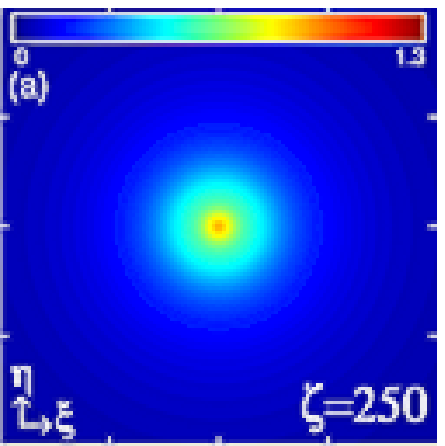}\includegraphics[width=0.4\linewidth]{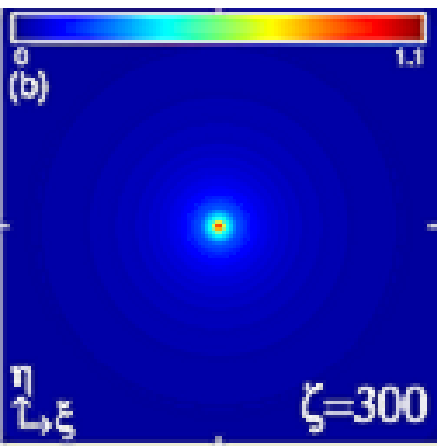}
\includegraphics[width=0.8\linewidth]{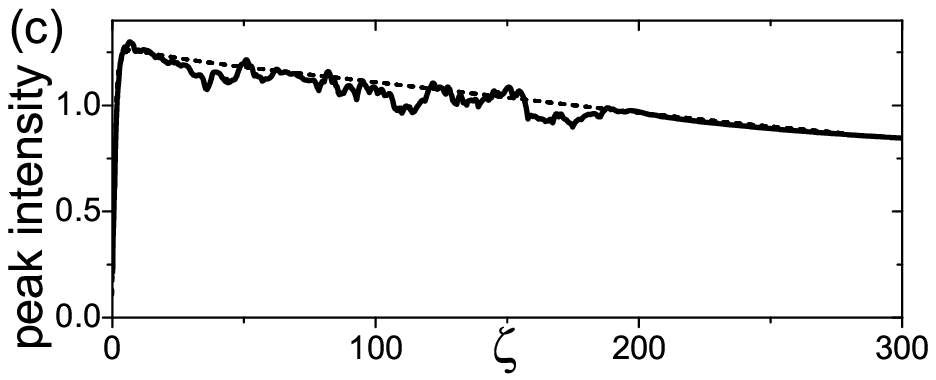}
\includegraphics[width=0.8\linewidth]{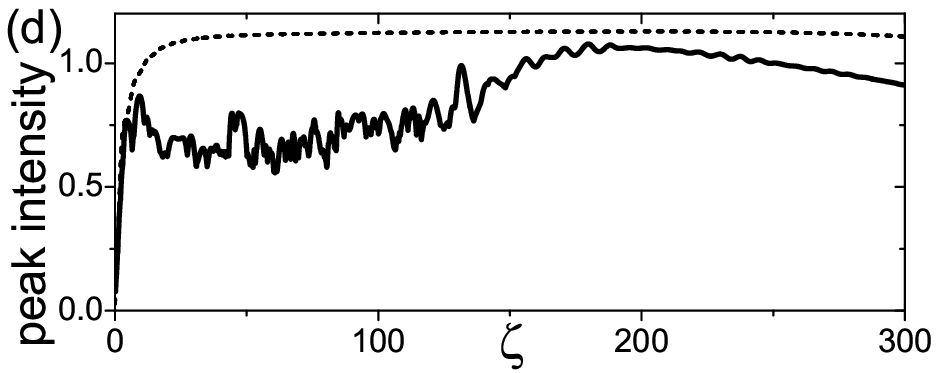}
\caption{\label{Fig10} (a) and (b) The same as in the lower panels in Figs. \ref{Fig9}(a) and \ref{Fig9}(b) but when the axicon is illuminated with clean Gaussian beams. (c) and (d) Solid curves: Peak intensity versus propagation distance for the numerical simulations in Fig. \ref{Fig9}(a) and Fig. \ref{Fig9}(b). Dashed curves: Peak intensity without nested vortices.}
\end{figure}

Vortex dipole annihilation in the NBB of the nonlinearly absorbing medium can be used massively to force the annihilation of many vortices of unit opposite charges acquired by speckled or scintillated light beams after propagation through random media such as a rough transparent plate or turbulent gaseous media \cite{VLADIMIROVA}. Considerable effort has been paid to reduce the speckle of scintillated beams using different techniques \cite{WANG}. A recent technique, somehow connected to the present work, is the use of adaptive optics to introduce a phase background that forces dipole annihilation \cite{RUX,RUX2}. Here, the appropriate phase background is the permanently converging wave front of the NBBs in the nonlinearly absorbing medium. We propose to transform the deteriorated beam into a conical beam by means of an axicon, or equivalent device, and let the conical beam propagate through a nonlinear absorbing medium placed immediately after the axicon, where the vortices will annihilate by pairs.

As explained at the end of Sec. \ref{NBB}, with a clean Gaussian beam $u=b_G \exp{(-\rho^2/\rho_0^2)}$ illuminating the axicon, a physically realizable, fundamental NBB ($n=0$) with finite power, of the cone angle imprinted by the axicon and with $C_{0}$ such that $|b_{\rm in}|^2=b_B^2 = b_G^2 \pi \rho_0\sqrt{2/e}$ will be formed about the middle of the Bessel zone of length $\zeta_B=\rho_0/\sqrt{2}$. With the deteriorated Gaussian beam with large number of randomly nested vortices of charges $+1$ and $-1$ and of total charge $s=0$ illuminating the axicon, the same NBB is expected to be formed, at the same time that vortices are pushed towards the NBB center and annhilate.

The top panels in Figs. \ref{Fig9}(a) and \ref{Fig9}(b) show central portions of the intensity pattern of such deteriorated Gaussian beams. They have been obtained by nesting the punctual vortices randomly in a broad, collimated Gaussian beam and letting the Gaussian beam with the vortices propagate in free space (linearly) a sufficiently long distance. If these deteriorated Gaussian beams illuminate an axicon, and self-defocusing [Fig. \ref{Fig9}(a)] or self-focusing [Fig. \ref{Fig9}(b)] media with significant nonlinear absorption are placed immediately after the axicon, intensity patterns at selected propagation distances within the Bessel zone are represented in the lower panels. Since the line-focus that will form the NBB in the Bessel zone is much more intense and narrower than the Gaussian beam, the color scale is changed and only the portions within the dashed rectangles of the top panels are shown. As expected, all vortices are heading towards the beam center, where they annihilate. The bottom panels in Figs. \ref{Fig9}(a) and \ref{Fig9}(b) about the end of the Bessel zone evidence that the results is a clean, vortex-less beam.  %The massive annihilation of the vortices can better be appreciated in the two videos in the supplemental material \cite{SUPP}.

Figure \ref{Fig10} supports that the massive vortex annihilation is due to the vortex attraction property of the finite-power NBB that tends to be formed, though constantly and severely perturbed, in the Bessel zone. The intensity patterns in the bottom panels of Figs. \ref{Fig9}(a) and \ref{Fig9}(b) at distances about the end of the Bessel zone closely resemble the intensity patterns in Figs. \ref{Fig10}(a) and \ref{Fig10}(b) at the same distances when a clean Gaussian beam illuminates the axicon. The residual deformation is due to the late annihilation of the vortices near the end of the Bessel zone, where the conical power flux from the axicon starts to be inefficient to maintain the NBB. Also, Figs. \ref{Fig10}(c) and \ref{Fig10}(d) show similar behaviors of the axial variation of the peak intensity when the axicon is illuminated by a clean Gaussian beam (dashed curves) and with the deteriorated Gaussian beam (solid curves). Deviations are more important with the NBB in self-focusing media because the collisions of the vortices constantly destroy the quite narrow central spike of the NBB, which nevertheless emerges when all vortices are washed out.

\section{Conclusions}

We have studied the dynamics of optical vortices nested in the diffraction-free, fundamental or high-order, nonlinear Bessel beams supported by self-focusing or self-defocusing media at intensities at which nonlinear absorption is significant. The behaviour of vortices contrasts strongly with  the behavior in standard Gaussian-like beams. In addition to the robustness of the vortex of arbitrary charge in the center of high-order nonlinear Bessel beams, the most relevant phenomenon that comes out from our analysis is the vortex attraction property introduced by the effect of the nonlinear absorption. Stationary propagation in media with nonlinear absorption requires permanently converging wave fronts, which according to well-established rules of vortex dynamics in a background beam, push any vortex placed anywhere towards the beam center. This property provides an efficient tool for optical vortex trapping, vortex combination and vortex dipole annihilation.\\

This mechanism is distinctive of nonlinear Bessel beams in nonlinearly absorbing media, and should be distinguished from the well-known self-healing property of linear or nonlinear Bessel beams. A linear Bessel beam, for example, self-heals after obstacles, but does not attract vortices. Also, vortex trapping in nonlinearly absorbing media is accompanied by a mode conversion and therefore by a permanent modification of the nonlinear Bessel beam.\\

We have explored the combination of two vortices into a single vortex, but the same mechanism can provide a $N$ to $1$ vortex combiner, which in turn can combine or mix the waves or the particles that the vortices can guide. Also, the vortex annihilation capability could be relevant for cleaning speckled or scintillated beams, a topic that has been extensively discussed \cite{RUX,VLADIMIROVA,WANG,RUX2}, and where complex solutions requiring substantial engineering have been described. We have proposed a simple setup able to remove undesired vortices without complicated step-by-step procedures or tight alignment requirements.\\

The author acknowledges support from Projects of the Spanish Ministerio de Econom\'{\i}a y Competitividad No. MTM2015-63914-P and No. FIS2017-87360-P.

\end{document}